
\magnification\magstep1
\font\twelverm=cmr10
\font\larbf=cmbx10 scaled \magstep1
\font\smit=cmti9
\font\smrm=cmr9
\baselineskip 0.6cm

\twelverm
\noindent{\larbf Chiral symmetry and the constituent quark model:}

\noindent{\larbf A null-plane point of view}
\medskip
{}~~~~~D. Mustaki

{}~~~~~{\smit Department of Physics and Astronomy,
Bowling Green State University,}

{}~~~~~{\smit Bowling Green, OH 43403}
\bigskip\noindent{\smrm
In order to clarify the connection between {\it current} and {\it constituent}
quarks (of u,d,s flavors), several authors have used the lightlike chiral
$SU(3)\otimes SU(3)$ algebra as a central concept. This literature is reviewed
here with the goal of offering an introduction to the subject within a
convenient, unified framework. It is shown that the null-plane Hamiltonian
for free massive fermions is chirally symmetric, provided only that the
particles have equal masses ($SU(3)$ limit). In the free quark model, hadrons
can be classified by a chiral $SU(3)\otimes SU(3)$ algebra, the generators of
which are current lightlike charges. Naturally, QCD interactions break chiral
symmetry, and the axial nonsinglet charges are not conserved. This remains
true in the 'chiral limit' (zero quark masses), signalling the spontaneous
breakdown of chiral symmetry. The actual generators of the $SU(3)\otimes
SU(3)$ classification of physical hadrons are obtained from the current
charges by means of a unitary transformation. Under this transformation,
quarks of the same flavor and of opposite helicities mix to form a constituent
quark. The functional form of this unitary transformation can be strongly
constrained on the basis of symmetry arguments. This analysis is potentially
rich in phenomenological applications, and a few are presented here. The
author offers also some new results: 1) it is shown that the null-plane axial
currents are different from their space-time counterparts, even in the $SU(3)$
limit; 2) when defining charges, the effect of boundary terms at infinity in
the longitudinal direction is taken into consideration. No prior knowledge of
light-cone formalism on the part of the reader is assumed.}

\null

\noindent{\bf CONTENTS}

\noindent~~I. Introduction

\noindent~II. Free Fermions

\noindent~~~~~A. Generalities

\noindent~~~~~B. Chiral Symmetries

\noindent~~~~~C. Chiral Charge and Fermion Number

\noindent~~~~~D. Flavor Symmetries

\noindent~~~~~E. Lightlike Chiral Algebra

\noindent~~~~~F. Conclusions

\noindent III. Quantum Chromodynamics

\noindent~~~~~A. Explicit Breaking of Chiral Symmetry

\noindent~~~~~B. Dynamical Breaking of Chiral Symmetry

\noindent~~~~~C. Physical Multiplets

\noindent~IV. Summary and Outlook

\noindent Acknowledgments

\noindent Appendix: Boundary Term for Free-Fermion Number

\noindent References

\null

\noindent {\bf I. INTRODUCTION}

``When you think about light hadrons, what picture comes naturally to your
mind?'' The answer to this question may be quite different from one
physicist to another depending on his/her subspecialty. Indeed we have
been using for many years two successful, but quite distinct, pictures
of hadronic structure: Quantum Chromodynamics (QCD), and the Constituent
Quark Model (CQM).

QCD emerged in the mid-70's from Current Algebra (which gave
birth to modern Chiral Perturbation Theory) on one hand, and from
the Parton Model (itself a product of the SLAC experiments
of the late 60's) on the other. In Current Algebra, one makes use of
the Partially
Conserved Axial-Current hypothesis (PCAC), which states that light hadrons
would be subjected to a fermionic symmetry called 'chiral
symmetry' if only the pion mass was zero. If this were the case, the
symmetry would be spontaneously broken, and the pions and kaons would be the
corresponding Goldstone bosons. As seen by PCAC, the real world slightly
misses this state of affairs by effects quantifiable in terms of the
pion mass and decay constant. This violation can be expressed in terms
of explicit symmetry breaking due to the nonzero masses of the fundamental
fermion fields (quarks of three light flavors),
and typically one assigns values of 4 MeV for the up-quark, 7MeV
for the down-quark and 130 MeV for the strange-quark (Gasser and Leutwyler,
1975; Shifman {\it et al.}, 1979). In a related
approach, the MIT bag model, one neglects the masses of the up- and
down-quarks, and uses a value of about 280 MeV for the strange-quark
(Hasenfratz and Kuti, 1978).

In the other picture, the Constituent Quark Model
(which can be traced back to the Eightfold Way), one always describes
mesons as made of a quark and an anti-quark, and baryons as made of three
quarks (or three anti-quarks).
These constituents are bound by some phenomenological potential which is
tuned to account for hadrons' properties such as masses, decay rates
or magnetic moments.

Unfortunately, there exist severe contradictions between these two
pictures:

\noindent\enskip $\bullet$ The QCD vacuum is an infinite sea of quark-antiquark
pairs; since the nonrelativistic Constituent Model ascribes low momenta
to the quarks,
these constituents should be strongly coupled to the sea quarks; but
then how can one identify two, or three, valence quarks in the sea?

\noindent\enskip $\bullet$ Since the coupling is strong, one expects that
vacuum polarization effects should be important; but the Constituent
Model cannot give account of these effects since there are no gluons,
and the number of constituents is fixed.

\noindent\enskip $\bullet$ The Constituent Quark Model does not display any
visible manifestation of spontaneous chiral symmetry breaking; actually,
it totally prohibits such a symmetry
since the constituent masses are large on a hadronic scale,
typically of the order of one-half of a meson mass or one-third of a baryon
mass; standard values are 330 MeV for the up- and down-quark, and
490 MeV for the strange-quark
(Georgi, 1982), very far from the 'current' masses
quoted above (even the ratio of the up- or down-quark mass to the strange-quark
mass is vastly different).

\noindent\enskip $\bullet$ If one attempted to
incorporate a bound gluon into the model, one would have
to assign to it a mass at least
of the order of magnitude of the quark mass in order to limit
its impact on the classification scheme;
but a gluon mass violates the gauge-invariance of QCD.

This long-standing incompatibility cannot be resolved in a
nonrelativistic framework. This observation has provided motivation for
a variety of relativistic CQM's, which, at least in the case of light quark
systems, certainly bring us closer to the phenomenology of QCD than the
nonrelativistic models. Nonetheless, all of these models take as
{\it ansatz} the existence of constituent quarks whose masses, after fitting
to the observed spectrum, turn out to be much larger than current masses.

It is however possible to relate in a natural way
the relativistic Constituent Quark Model to the underlying
field theory, provided one sits
in a {\it null-plane frame} rather than in space-time. We shall elaborate in
great detail about why this choice of metric is key to exposing the
connection we are seeking between current and constituent
quarks\footnote{$^1$}{No prior
knowledge of light-cone formalism on the part of the reader is
assumed, and all necessary definitions are provided in the text.}.

The idea of deriving a Null-Plane Constituent Model from QCD actually
dates from
the early seventies, and there is a rich literature on the subject
(Buccella {\it et al.}, 1970; De Alwis, 1973; Eichten {\it et al.}, 1973;
Bell, 1974; De Alwis and Stern, 1974; Leutwyler, 1974b, 1974c;
Melosh, 1974; Osborn, 1974; Carlitz {\it et al.}, 1975;
Ida, 1975b, 1975c; Carlitz and Tung, 1976).
The purpose of this article is to review in varying degrees of detail
the theories that have been proposed, with the goal
of presenting a self-consistent framework rather than trying to cover
the subject exhaustively. Along
the way, we clarify some obscure or little-known aspects, and offer some new
results. Among the latter, we will show that the space-time and null-plane
axial currents are distinct; this remark is at the root of the difference
between the chiral properties of QCD in the two frames.

The emphasis of the present work will be on concepts rather than on
calculational techniques, with special attention devoted to the contrasting
implications of flavor symmetries in the two frames.
Lengthy derivations will be offered only in a few cases where
(to our knowledge) no proof has been provided in the literature.
Otherwise, we will guide the interested reader to the relevant sources.

The organization of the article is a follows. In part II we present
the vector and axial properties of free fermions, first for
a single flavor, then in the framework of approximate $SU(3)$ symmetry.
In part III we discuss how QCD behaves under chiral flavor transformations,
progressively working our way to the properties of real-world hadrons. The
connection between current and constituent quarks is finally elucidated. In
part IV, we summarize our results, and discuss a few unresolved issues.

\null

\noindent{\bf II. FREE FERMIONS}

\null

\noindent{\bf A. Generalities}

	In our metric, we use
$$
x^{\pm} \equiv {x^0 \pm x^3 \over \sqrt{2}}~~,
$$
where $x^+$ is the 'light-cone time' and $x^-$ is the 'longitudinal'
coordinate.

Consider to begin with the free theory of fermions of a single flavor. From
the Lagrangian density
$$
{\cal L} =\bar\psi ({i\over 2}\buildrel\leftrightarrow\over{\not\!{\partial}}
-m)\psi~~, \eqno (1)
$$
one derives the Dirac equation
$$
(i\not\!{\partial}-m)\psi=0~~,~~\bar\psi
(i\buildrel \leftarrow \over {\not\!{\partial}} +m)=0~~, \eqno (2)
$$
and the energy-momentum tensor
$$
T^{\mu\nu}=\bar\psi \Biggl[ {i\over 2}\gamma^{\mu}
\buildrel\leftrightarrow\over{\partial^{\nu}} +g^{\mu\nu}(m-{i\over 2}
\buildrel\leftrightarrow\over{\not\!{\partial}} )\Biggr]\psi~~. \eqno (3)
$$

In a space-time frame, the energy-momentum operator is
$$
P^{\mu}=\int d^3 {\bf x}~T^{\mu 0}~~. \eqno (4)
$$
In particular, the Hamiltonian is
$$
P^0 =\int d^3 {\bf x}~T^{00}~~, \eqno (5)
$$
where
$$
T^{00}=\bar\psi (-i\vec\gamma \cdot\vec\partial +m)\psi~~. \eqno (6)
$$

In a null-plane frame, the 'energy-momentum' operator is
$$
P^{\mu}=\int d^3 \tilde{x}~T^{\mu +}~~, \eqno (7)
$$
where
$$
d^3 \tilde{x}\equiv dx^- \,d^2 {\bf x}_{\perp}~~. \eqno (8)
$$
In particular, the energy operator is
$$
P^{-}=\int d^3 \tilde{x}~T^{-+}~~, \eqno (9)
$$
where
$$
T^{-+}={i\over 2}\,\bar\psi \gamma^- \buildrel\leftrightarrow\over
{\partial_-} \psi ~~. \eqno (10)
$$
The matrices
$$
\Lambda_{\pm}\equiv {\gamma ^{\mp}\gamma ^{\pm} \over 2}~~, \eqno (11)
$$
where
$$
\gamma^{\pm}\equiv{\gamma^0 \pm\gamma^3 \over\sqrt{2}}~~, \eqno (12)
$$
are Hermitian projection operators, viz.,
$$
(\Lambda_{\pm})^{2}=\Lambda_{\pm}~~,~~\Lambda_{\pm}\Lambda_{\mp}=0~~,~~
\Lambda_+ +\Lambda_- ={\bf 1}~~~. \eqno (13)
$$
Their action on Dirac spinors yields
$$
\psi _{\pm}\equiv \Lambda_{\pm}\psi~~~. \eqno (14)
$$
The reason for splitting the Dirac field $\psi$ into ($\psi_{+} +
\psi_{-}$) is that the Dirac equation shows that $\psi_{-}$ is a
dependent field. Indeed with, say, antiperiodic boundary conditions at
$x^{-}$ infinity, Eq. (2) can be solved for
$$
\psi_{-} (x)=-{i\over 4}\int dy^- \,\epsilon(x^- -y^-)\,(i\vec\gamma
_{\perp}\cdot\vec{\partial} _{\perp}+m)\gamma^+ \psi_+ (y)~~, \eqno (15)
$$
where~~${\bf y}_{\perp}\!=\!{\bf x}_{\perp}~$.
This means that the physical fermion states are built out of $\psi_{+}$.
Substituting Eq. (15) into Eqs. (9) and (10), one finds after an
integration by parts
$$
P^{-}={i\sqrt{2}\over 4}\int d^3 \tilde{x}\int dy^- \epsilon(x^-
-y^-)\,\psi^{\dagger}_+ (y)\,(m^2 -\Delta_{\perp})
\psi_+ (x)~~. \eqno (16)
$$

To any given transformation of the fermion field we associate a
current
$$
{\delta\cal L \over\delta(\partial_{\mu}\psi)}{\delta\psi\over\theta}
=i\bar\psi \gamma^\mu {\delta\psi \over \theta}~~, \eqno (17)
$$
where $\delta\psi$ is the infinitesimal variation parametrized
by $\theta$. For example the vector transformation is defined in
space-time by
$$
\psi\mapsto e^{-i\theta} \psi~~,~~\delta\psi =-i\theta\psi~~, \eqno (18)
$$
whence the current
$$
j^{\mu} =\bar\psi \gamma^{\mu} \psi~~. \eqno (19)
$$

In a null-plane frame, in view of the constraints structure, the vector
transformation will be defined as
$$
\psi_{+}\mapsto e^{-i\theta} \psi_{+}~~,~~\delta\psi_{+} =-i\theta\psi_{+}
{}~~,~~\delta\psi=\delta\psi_{+} +\delta\psi_{-}~~, \eqno (20)
$$
where $\delta\psi_{-}$ is calculated using Eq. (15). The distinction
in the case of vector U(1) is of course academic:
$$
\delta\psi_{-}=-i\theta\psi_{-}~~\Longrightarrow~~\delta\psi=
-i\theta\psi~~, \eqno (21)
$$
therefore
$$
\tilde{j} ^{\mu}=j^{\mu}~~. \eqno (22)
$$
In contrast, we shall see in the next section that the axial-vector
currents (defined respectively in a space-time and in a null-plane
frame) are {\it not} equal. Finally, using Eq. (2), one checks easily
that the vector current is conserved:
$$
\partial_{\mu}j^{\mu}=0~~, \eqno (23)
$$
therefore the space-time and null-plane vector charges (which measure
fermion number)
$$
Q\equiv\int d^3 {\bf x}~j^0 (x)~~,~~\tilde{Q}\equiv\int d^3 \tilde{x}~
j^+ (x) \eqno (24)
$$
are equal (McCartor, 1988) (we show in Appendix A that the boundary
term at $x^-$-infinity vanishes).

\null

\noindent{\bf B. Chiral symmetries}

The space-time chiral transformation is defined by
$$
\psi\mapsto e^{-i\theta\gamma_5} \psi~~,~~\delta\psi=-i\theta
\gamma_5 \psi~~, \eqno (25)
$$
where $\gamma_5 \equiv i\gamma^{0}\gamma^{1}\gamma^{2}\gamma^{3}$. From
Eqs. (5) and (6), one sees that the space-time theory with nonzero
fermion masses is not chirally symmetric.

The null-plane chiral transformation is
$$
\psi_{+}\mapsto e^{-i\theta\gamma_5} \psi_{+}~~,~~\delta\psi_{+}=
-i\theta\gamma_5 \psi_{+}~~. \eqno (26)
$$
This {\it is} a symmetry of the null-plane theory, as seen from Eq. (16),
without requiring zero bare masses.

The space-time axial-vector current associated to the transformation
Eq. (25) is
$$
j^{\mu}_5 =\bar\psi \gamma^{\mu}\gamma_5 \psi ~~. \eqno (27)
$$
Using Eq. (2), one obtains
$$
\partial_{\mu}j^{\mu}_5 =2im\bar\psi \gamma_5 \psi ~~. \eqno (28)
$$
As expected, this current is not conserved for nonzero fermion mass. The
associated charge is
$$
Q_5 \equiv\int d^3 {\bf x}~ j^{0}_{5}=\int d^3 {\bf x}~\bar\psi \gamma^0
\gamma_5 \psi~~~. \eqno (29)
$$

Inserting Eq. (26) in Eq. (15), and using $\{\gamma^{\mu},\gamma_5\}=0$,
one finds
$$
\delta\psi_{-} (x)=-\theta\gamma_5 \int dy^- \,{\epsilon (x^- -y^- )
\over 4}(i\vec{\gamma}_{\perp}\cdot\vec{\partial}_{\perp}-m)
\gamma^+ \psi_+ (y)~~. \eqno (30)
$$
This expression differs from
$$
-i\theta\gamma_5 \psi_{-} =-\theta\gamma_5 \int dy^- \,{\epsilon (x^- -y^- )
\over 4}(i\vec{\gamma}_{\perp}\cdot\vec{\partial}_{\perp}+m)
\gamma^+ \psi_+ (y) \eqno (31)
$$
(again, for nonzero masses), therefore ${\tilde j}^{\mu}_5 \not= j^{\mu}_5$
(except for the plus component, due to $(\gamma^+)^2 =0$). To be precise,
$$
{\tilde j}^{\mu}_5 = j^{\mu}_5 +im\bar\psi \gamma^{\mu} \gamma_5 \int
dy^- \,{\epsilon(x^- -y^- )\over 2}\gamma^+ \psi_+ (y)~~. \eqno (32)
$$
Using Eqs. (28) and (32), and the identity
$$
\not\!{\partial}\,\int dy^- ~{\epsilon (x^- -y^- )\over 4}\,\gamma^+ \psi_+ (y)
=\psi_+ (x)+\int dy^- ~{\epsilon (x^- -y^- )\over 4}\,
\vec{\gamma}_{\perp}\cdot\vec{\partial}_{\perp}\,\gamma^+ \psi_+ (y)~~,
\eqno (33)
$$
a straightforward calculation shows that
$$
\partial_{\mu} {\tilde j} ^{\mu}_5 =0 \eqno (34)
$$
as expected. Finally the null-plane chiral charge is
$$
{\tilde Q}_5 \equiv\int d^3 {\tilde x}~ {\tilde j}^{+}_{5}=\int d^3 {\tilde x}
{}~\bar\psi \gamma^{+} \gamma_5 \psi \eqno (35)
$$
(one can write here $\psi$ or $\psi_{+}$ indifferently due to the
$\gamma^{+}$ factor).
\vskip 1.2cm

\noindent{\bf C. Chiral charge and fermion number}

{}From the canonical anti-commutator
$$
\{\psi(x),\psi^{\dagger}(y)\}_{x^0 =y^0}=\delta^3 ({\bf x}-{\bf y})~~,
\eqno (36)
$$
one derives
$$
[\psi,Q_5 ]=\gamma_5 \psi~~\Longrightarrow~~[Q,Q_5]=0~~, \eqno (37)
$$
so that fermion number, viz., the number of
quarks minus the number of anti-quarks is conserved by the chiral charge.
However, the latter are not conserved {\it separately}. This
can be seen by using the momentum expansion of the field
$$
\psi(x)=\int {d^3 {\bf p}\over (2\pi)^{3/2} 2p^0}\sum_{s=\pm 1}
\Biggl[u({\bf p},s)e^{-ipx}b({\bf p},s)
+v(-{\bf p},-s)e^{+ipx}d^{\dagger}({\bf p},s)\Biggr]~~,\eqno (38)
$$
where $~px\equiv p^0 x^0 -{\bf p}\cdot{\bf x}~,~p^0 =\sqrt{{\bf p}^2
+m^2}~$, and
$$
\{b({\bf p},s),b^{\dagger}({\bf q},s')\}=2p^0 \delta^3 ({\bf p}-{\bf q})
\delta_{ss'}=\{d({\bf p},s),d^{\dagger}({\bf q},s')\}~~, \eqno (39)
$$
$$
\sum_{s=\pm 1}u({\bf p},s)\bar u ({\bf p},s)=\not\!p +m~,~
\sum_{s=\pm 1}v(-{\bf p},s)\bar v ({-\bf p},s)=\not\!p -m~~. \eqno (40)
$$
Inserting Eq. (38) in Eq. (29) yields indeed
$$\eqalign{
Q_5 =\int {d^3 {\bf p}\over 2p^0} \sum_{s=\pm 1}s &\Biggl[
{|{\bf p}|\over p^0}\biggl( b^{\dagger}({\bf p},s)b({\bf p},s)+d^{\dagger}
({\bf p},s)d({\bf p},s)\biggr) \cr
&+{m\over p^0}\biggl( d^{\dagger}(-{\bf p},s)b^{\dagger}({\bf p},s)
e^{2ip^0 t}+b({\bf p},s)d(-{\bf p},s)e^{-2ip^0 t}\biggr)\Biggr]~~.\cr}
\eqno (41)
$$
This implies that when $Q_5$ acts on a hadronic state, it will add
or absorb a continuum of
quark-antiquark pairs (the well-known pion pole) with a probability
amplitude proportional to the fermion mass and inversely proportional
to the energy of the pair. Because of that, $Q_5$ is most unsuited
for classification purposes.

In contrast, the null-plane chiral charge conserves not only fermion
number (electric charge), but also the number of
quarks and anti-quarks separately. In effect, the canonical anti-commutator
is
$$
\{\psi_+ (x),\psi^{\dagger}_+ (y)\}_{x^+ =y^+}={\Lambda_+ \over\sqrt{2}}
\,\delta^3 ({\tilde x}-{\tilde y})~~,\eqno (42)
$$
hence the momentum expansion of the field reads
$$
\psi_{+}(x)=\int {d^3 \tilde{p} \over (2\pi)^{3/2} 2^{3/4} \sqrt{p^{+}}}
\sum_{h=\pm {1\over 2}}\Biggl[w(h)e^{-ipx}b(\tilde{p},h)
+w(-h)e^{+ipx}d^{\dagger}(\tilde{p},h)\Biggr]~~,\eqno (43)
$$
where $~px\equiv p^{-}x^{+}+p^{+}x^{-}-{\bf p}_{\perp}\cdot
{\bf x}_{\perp}~,~p^- ={{\bf p}_{\perp}^2 +m^2 \over 2p^+}~$, and
$$
\{b(\tilde{p},h),b^{\dagger}(\tilde{q},h')\}=2p^{+}\delta^3 (\tilde{p}-
\tilde{q})\delta_{hh'}=\{d(\tilde{p},h),d^{\dagger}(\tilde{q},h')\}~~,
\eqno (44)
$$
$$
\sum_{h=\pm{1\over 2}}w(h)w^{\dagger}(h)=\Lambda_+ ~~.\eqno (45)
$$
(In the rest frame of a system, its total angular momentum along the $z$-axis
is called 'null-plane helicity'; the helicity of an elementary particle is
just the usual
spin projection; we label the eigenvalues of helicity
with the letter 'h'.) It is easiest to work in the so-called
'chiral representation' of Dirac matrices, where
$$
\gamma_5 =\left[\matrix{
1&0&0&0\cr
0&1&0&0\cr
0&0&-1&0\cr
0&0&0&-1\cr}\right]~,~w(+{1\over 2})=\left[\matrix{1\cr 0 \cr 0\cr 0\cr}
\right]~,~w(-{1\over 2})=\left[\matrix{0\cr 0\cr 0 \cr 1\cr}\right]
$$
$$
\Longrightarrow ~w^{\dagger}(h)\gamma_5 w(h')=2h\delta_{hh'}~~.\eqno (46)
$$
Inserting Eq. (43) into Eq. (35), one finds
$$
\tilde{Q}_5 =\int {d^3 \tilde{p}\over 2p^{+}}\sum_h 2h\, \Biggl[ b^{\dagger}
(\tilde{p},h)b(\tilde{p},h)+d^{\dagger}(\tilde{p},h)d(\tilde{p},h)\Biggr]~~.
\eqno (47)
$$
This is just a superposition of fermion and anti-fermion
number operators, and thus our claim is proved.
This expression also shows that $\tilde{Q}_5$ annihilates the vacuum,
and that it simply measures (twice) the sum
of the helicities of all the quarks and anti-quarks ('constituents') of
a given state. Indeed, in a null-plane frame, the handedness of an
individual fermion is automatically determined by its helicity. To show
this, note that
$$
\gamma_5 w(\pm {1\over 2})=\pm w(\pm {1\over 2})~~\Longrightarrow ~~
{1\pm\gamma_5 \over 2} w(\pm{1\over 2})=w(\pm{1\over 2})~,~{1\pm\gamma_5
\over 2}w(\mp{1\over 2})=0~~. \eqno (48)
$$
Defining as usual
$$
\psi_{+R}\equiv {1+\gamma_5 \over 2}\psi_{+}~,~\psi_{+L}\equiv {1-\gamma_5
\over 2}\psi_{+}~~,\eqno (49)
$$
it follows from Eqs. (43) and (48) that $\psi_{+R}$ contains only
fermions of helicity $+{1\over 2}$ and anti-fermions of helicity
$-{1\over 2}$, while $\psi_{+L}$ contains only fermions of helicity
$-{1\over 2}$ and anti-fermions of helicity $+{1\over 2}$.
Also, we see that when acted upon by the right- and left-hand charges
$$
\tilde{Q}_{R} \equiv {\tilde{Q} +\tilde{Q}_5 \over 2}~,~\tilde{Q}_{L} \equiv
{\tilde{Q} -\tilde{Q}_5 \over 2}~~,\eqno (50)
$$
a chiral fermion (resp. anti-fermion) state may have eigenvalues
$+1$ (resp. $-1$) or zero.

In a space-time frame, this identification between helicity and chirality
applies only to massless fermions.

\null

\noindent{\bf D. Flavor symmetries}

We proceed now to the theory of three flavors of free fermions $\psi_f$,
where $f=u,d,s$. Based on Eqs. (5) and (6), the space-time Hamiltonian is
$$\eqalign{
P^0 &=\sum_f \int d^3 {\bf x}~\bar\psi_f \,(-i\vec\gamma \cdot\vec\partial
+m_f)\,\psi_f \cr
&=\int d^3 {\bf x}~\bar\psi\,(-i\vec\gamma \cdot\vec\partial +M)\,\psi~~,}
\eqno (51)
$$
where now
$$
\psi\equiv\left[\matrix{\psi_u \cr \psi_d \cr \psi_s \cr}\right]
{}~,~{\rm and}~~M\equiv\left[\matrix{
m_u & 0 & 0 \cr
0 & m_d & 0 \cr
0 & 0 & m_s \cr
}\right]~~.
$$
The vector, and axial-vector, flavor nonsinglet transformations are defined
respectively as
$$
\psi\mapsto e^{-i{\lambda^{\alpha}\over 2}\theta^{\alpha}} \psi~~,~~\psi
\mapsto e^{-i{\lambda^{\alpha}\over 2}\theta^{\alpha}\gamma_5}\psi~~,\eqno (52)
$$
where the summation index $\alpha$ runs from 1 to 8.
$P^0$ is invariant under vector transformations if the quarks have equal
masses ('$SU(3)$ limit'), and invariant under chiral transformations if all
masses are zero ('chiral limit').

{}From Eq. (16), the null-plane Hamiltonian is
$$\eqalign{
P^{-}&=\sum_f {i\sqrt{2}\over 4}\int d^3 \tilde{x}\int dy^- \,\epsilon(x^-
-y^-)\,\psi^{\dagger}_{f+} (y)\,(m_f ^2 -\Delta_{\perp})\psi_{f+} (x)\cr
&={i\sqrt{2}\over 4}\int d^3 \tilde{x}\int dy^- \,\epsilon(x^-
-y^-)\,\psi^{\dagger}_{+} (y)\,(M^2 -\Delta_{\perp})\psi_{+} (x)~~.}\eqno (53)
$$
Naturally,  $P^-$ is not invariant under the vector transformations
$$
\psi_+ \mapsto e^{-i{\lambda^{\alpha}\over 2}\theta^{\alpha}} \psi_{+}
\eqno (54)
$$
unless the quarks have equal masses. But it they do, then
$P^-$ is also invariant under the chiral transformations
$$
\psi_+ \mapsto e^{-i{\lambda^{\alpha}\over 2}\theta^{\alpha}\gamma_5}
\psi_{+}~~, \eqno (55)
$$
whether this common mass is zero or not.

Writing the Dirac equation for the three flavors in compact form:
$$
(i\not\!{\partial}-M)\psi=0~~,~~\bar\psi
(i\buildrel \leftarrow \over {\not\!{\partial}} +M)=0~~, \eqno (56)
$$
one finds that the space-time currents
$$
j^{\mu\alpha} =\bar\psi \gamma^{\mu} {\lambda^{\alpha}\over 2}\psi~~,~~
j^{\mu\alpha}_5 =\bar\psi \gamma^{\mu}\gamma_5 {\lambda^{\alpha}\over 2}
\psi ~~, \eqno (57)
$$
have the following divergences:
$$
\partial_{\mu}j^{\mu\alpha} =i\bar\psi \biggl[M,{\lambda^{\alpha}\over 2}
\biggr]\psi~~,~~\partial_{\mu}j^{\mu\alpha}_5 =i\bar\psi \gamma_5
\biggl\{ M,{\lambda^{\alpha}\over 2}\biggr\}\psi~~. \eqno (58)
$$
These currents have obviously the expected conservation properties.

Turning to the null-plane frame, we rewrite Eq. (15) as
$$
\psi_- (x)=-{i\over 4}\int dy^- \,\epsilon(x^- -y^-)\,(i\vec\gamma
_{\perp}\cdot\vec{\partial} _{\perp}+M)\gamma^+ \psi_+ (y)~~. \eqno (59)
$$
Then we find
$$
\tilde{j}^{\mu\alpha}=j^{\mu\alpha}-i\bar\psi \biggl[ M,{\lambda^{\alpha}
\over 2}\biggr]\gamma^{\mu}\int dy^- \,{\epsilon(x^- -y^-)\over 4}\,
\gamma^+ \psi_+ (y)~~. \eqno (60)
$$
So $\tilde{j}^{\mu\alpha}$ and $j^{\mu\alpha}$ may be equal for all $\mu$
only if the quarks have equal masses. The {\it vector}, flavor nonsinglet
charges in each frame are two different octets of operators,
except in the $SU(3)$ limit.

For the null-plane current associated with axial transformations, we get
$$
\tilde{j}^{\mu\alpha}_5 =j^{\mu\alpha}_5 -i\bar\psi \biggl\{ M,
{\lambda^{\alpha}\over 2}\biggr\}\gamma_5 \gamma^{\mu}\int dy^- \,
{\epsilon(x^- -y^-)\over 4}\,\gamma^+ \psi_+ (y)~~. \eqno (61)
$$
Hence $\tilde{j}^{\mu\alpha}_5$ and $j^{\mu\alpha}_5$ are not equal
(except for $\mu =+$), even in the $SU(3)$ limit, unless all quark masses
are zero. Finally, one obtains the following divergences:
$$\eqalign{
\partial_{\mu}\tilde{j}^{\mu\alpha}&=\bar\psi \biggl[ M^2 ,{\lambda^{\alpha}
\over 2}\biggr]\int dy^- \,{\epsilon(x^- -y^-)\over 4}\,\gamma^+
\psi_+ (y)~~,\cr
\partial_{\mu}\tilde{j}^{\mu\alpha}_5 &=-\bar\psi \biggl[ M^2 ,
{\lambda^{\alpha}\over 2}\biggr]\gamma_5 \int dy^- \,{\epsilon(x^- -y^-)
\over 4}\,\gamma^+ \psi_+ (y)~~.} \eqno (62)
$$
As expected, both null-plane currents are conserved in the $SU(3)$ limit,
without requiring zero masses. Also note how null-plane relations often seem
to involve
the masses {\it squared}, while the corresponding space-time relations are
{\it linear} in the masses. The integral operator
$$
\int dy^- \,{\epsilon(x^- -y^-)\over 2}~\equiv~{1\over \partial^{x}_-}
$$
compensates for the extra power of mass.

\null

\noindent{\bf E. Lightlike chiral algebra}

The associated null-plane charges are
$$
\tilde {Q}^{\alpha} \equiv \int d^3 {\tilde x} \,\bar{\psi} \gamma^{+}
{\lambda^{\alpha}\over 2}\psi~,~
\tilde {Q}^{\alpha}_5  \equiv \int d^3 {\tilde x} \,\bar{\psi} \gamma^{+}
\gamma_5 {\lambda^{\alpha}\over 2}\psi~~. \eqno (63)
$$
Using the momentum expansion of the fermion triplet Eq. (43), where now
$$
b(\tilde{p},h)\equiv\left[\matrix{b_u (\tilde{p},h)\cr b_d (\tilde{p},h)
\cr b_s (\tilde{p},h)\cr}\right]~,~{\rm and}~~d(\tilde{p},h)\equiv\left[
\matrix{d_u (\tilde{p},h)\cr
d_d (\tilde{p},h)\cr d_s (\tilde{p},h)\cr}\right]~~,
$$
one can express the charges as
$$
\tilde{Q}^{\alpha} =\int {d^3 \tilde{p}\over 2p^{+}}\sum_h \Biggl[ b^{\dagger}
(\tilde{p},h){\lambda^{\alpha}\over 2}b(\tilde{p},h)-d^{\dagger}(\tilde{p},h)
{\lambda^{\alpha T}\over 2}d(\tilde{p},h)\Biggr]~~,
\eqno (64)
$$
$$\tilde{Q}^{\alpha}_5 =\int {d^3 \tilde{p}\over 2p^{+}}\sum_h 2h\,
\Biggl[ b^{\dagger}
(\tilde{p},h){\lambda^{\alpha}\over 2}b(\tilde{p},h)+d^{\dagger}(\tilde{p},h)
{\lambda^{\alpha T}\over 2}d(\tilde{p},h)\Biggr]~~,
\eqno (65)
$$
where the superscript $T$ denotes matrix transposition. Clearly all
sixteen charges annihilate the vacuum (Jers\'ak and Stern, 1968, 1969;
Leutwyler, 1968, 1969; Ida, 1975a; Sazdjian and Stern, 1975).

As $\tilde {Q}^{\alpha}$ and $\tilde {Q}^{\alpha}_5$ conserve the number
of quarks and anti-quarks separately, these charges are well-suited for
classifying hadrons in terms of their valence constituents, {\it whether the
quarks masses are equal or not} (De Alwis and Stern, 1974). Since the
charges commute with
$P^+$ and ${\bf P}_{\perp}$, all hadrons belonging to the same multiplet
have same momentum. But this common value of momentum is arbitrary, because
in a null-plane frame one can boost between any two values of momentum, using
only kinematic operators.

Using Eq. (42) and the $SU(3)$ commutation relations, one finds that
these charges generate an $SU(3)\otimes SU(3)$ algebra:
$$
[\tilde{Q}^{\alpha},\tilde{Q}^{\beta}]=i\,f_{\alpha\beta\gamma}
\,\tilde{Q}^{\gamma}~,~
[\tilde{Q}^{\alpha},\tilde{Q}^{\beta}_5 ]=i\,f_{\alpha\beta\gamma}
\,\tilde{Q}^{\gamma}_5 ~,~
[\tilde{Q}^{\alpha}_5 ,\tilde{Q}^{\beta}_5 ]=i\,f_{\alpha\beta\gamma}
\,\tilde{Q}^{\gamma}~~,\eqno (66)
$$
and the corresponding right- and left-hand charges generate two commuting
algebras denoted
$SU(3)_R$ and $SU(3)_L$ (Jers\'ak and Stern, 1968, 1969;
Leutwyler, 1968, 1969, 1974b, 1974c;
Buccella {\it et al.}, 1970; De Alwis, 1973; Eichten {\it et al.}, 1973;
Feinberg, 1973;
Bell, 1974; De Alwis and Stern, 1974; Ida, 1974, 1975a, 1975b, 1975c;
Melosh, 1974; Osborn, 1974; Carlitz {\it et al.}, 1975; Sazdjian and Stern,
1975; Carlitz and Tung, 1976)\footnote{$^1$}{Most of these papers in fact
study a larger algebra of lightlike charges, namely $SU(6)$, but
the subalgebra $SU(3)_R\otimes SU(3)_L$ suffices for our purposes.}.

Since
$$
[\psi_+ ,\tilde{Q}^{\alpha}_5 ]=\gamma_5 {\lambda^{\alpha} \over 2}
\psi_+ ~~,\eqno (67)
$$
the quarks form an irreducible representation of this algebra. To be
precise, the
quarks (resp. anti-quarks) with helicity $+{1\over 2}$ (resp. $-{1\over
2}$) transform as a triplet of $SU(3)_R$ and a singlet of $SU(3)_L$, the
quarks (resp. anti-quarks) with helicity $-{1\over 2}$ (resp. $+{1\over
2}$) transform as a triplet of $SU(3)_L$ and a singlet of $SU(3)_R$. Then
for example the ordinary vector $SU(3)$ decuplet of $J={3\over 2}$ baryons
with $h=+{3\over 2}$ is a pure right-handed (10,1) under $SU(3)_R\otimes
SU(3)_L$.
The octet ($J={1\over 2}$) and decuplet ($J={3\over 2}$) with $h=+{1\over 2}$
transform together as a (6,3). For bosonic states we expect both chiralities
to contribute with equal probability. For example, the
octet of pseudoscalar mesons arises from a superposition
of irreducible representations of $SU(3)_R\otimes SU(3)_L$:
$$
|J^{PC}=0^{-+}>={1\over \sqrt{2}}\,|(8,1)-(1,8)>~~,\eqno (68)
$$
while the octet of vector mesons with zero helicity corresponds to
$$
|J^{PC}=1^{--}>={1\over \sqrt{2}}\,|(8,1)+(1,8)>~~,\eqno (69)
$$
and so on. These low-lying states have $L_z =0$, where
$$
L_z=-i\int d^3 \tilde{x}~\bar\psi \gamma^+ (x^1 \partial_2 -x^2
\partial_1)\psi \eqno (70)
$$
is the orbital angular momentum along $z$.

In the realistic case of unequal masses, the chiral charges are not conserved.
Hence they generates multiplets which
are not mass-degenerate --- a welcome feature.
The fact that the invariance of the vacuum
does not enforce the 'invariance of the world' (viz., of energy), in sharp
contrast with the order of things in space-time (Coleman's theorem), is yet
another remarkable property of the null-plane frame.

\null

\noindent{\bf F. Conclusions}

In contrast with the space-time picture, free null-plane {\it current}
quarks are also {\it constituent} quarks because:

\noindent$\bullet$ They can be massive without preventing chiral
symmetry, which we know is (approximately) obeyed by hadrons.

\noindent$\bullet$ They form a basis for a classification of hadrons
under the lightlike chiral algebra.

\null

\noindent{\bf III. QUANTUM CHROMODYNAMICS}

\null

\noindent{\bf A. Explicit breaking of chiral symmetry}

In the quark-quark-gluon vertex $gj^{\mu}A_{\mu}$, the transverse
component of the vector current is
$$
{\bf j_{\perp}}(x)=\cdots +{im\over 4}\int dy^{-}\, \epsilon(x^{-}
-y^{-})\, \Biggl[\bar{\psi}_{+}(y)\gamma^{+}\vec{\gamma}_{\perp}
\psi_{+}(x)+\bar{\psi}_{+}(x)\gamma^{+}\vec{\gamma}_{\perp}\psi_{+}(y)
\Biggr]~~,\eqno(71)
$$
where the dots represent chirally symmetric terms, and where color, as well
as flavor, factors and indices have been omitted for clarity. The term
explicitly
written out breaks chiral symmetry for nonzero quark mass. Not
surprisingly, it generates
vertices in which the two quark lines have opposite
helicity.

The canonical anticommutator Eq. (42) for the bare fermion fields
still holds in the interactive theory (for each flavor). The momentum
expansion Eq. (43) gets rewritten as
$$
\psi_{+}(x)=\int {d^3 \tilde{p} \over (2\pi)^{3/2} 2^{3/4} \sqrt{p^{+}}}
\sum_{h=\pm {1\over 2}}\Biggl[w(h)e^{-i\tilde{p}\tilde{x}}b(\tilde{p},h,x^+)
+w(-h)e^{+i\tilde{p}\tilde{x}}d^{\dagger}(\tilde{p},h,x^+)\Biggr]~~,\eqno (72)
$$
where
$$
\tilde{p}\tilde{x} \equiv p^{+}x^{-}-{\bf p}_{\perp}\cdot
{\bf x}_{\perp}~~,\eqno (73)
$$
and
$$
\{b(\tilde{p},h,x^+),b^{\dagger}(\tilde{q},h',y^+)\}_{x^+ =y^+}
=2p^{+}\delta^3 (\tilde{p}-
\tilde{q})\delta_{hh'}=\{d(\tilde{p},h,x^+),d^{\dagger}(\tilde{q},h',y^+)\}
_{x^+ =y^+}~~. \eqno (74)
$$
The momentum expansions of the lightlike charges remain the same as in
Eqs. (64)-(65) (keeping in mind that the creation and annihilation
operators are now unknown functions of 'time'). Hence the charges still
annihilate the Fock vacuum, and are suitable for classification purposes.

We do not require annihilation of the physical vacuum (QCD
ground state). The successes of CQM's suggest that to understand the
properties of the hadronic spectrum, it may not be necessary to take
the physical vacuum into account. This is also the point of view taken
by the authors of a recent paper on the renormalization of QCD
(G{\l}azek {\it et al.}, 1994). Their
approach consists in imposing an 'infrared' cutoff in longitudinal
momentum, and in compensating for this suppression by means of Hamiltonian
counterterms. Now, only terms that annihilate the {\it Fock vacuum} are
allowed in their Hamiltonian $P^-$. Since all states in the truncated
Hilbert space have strictly positive longitudinal momentum except for the
Fock vacuum (which has $p^+ =0$), the authors hope to be able to adjust the
renormalizations in order to fit the observed spectrum, without having to
solve first for the physical vacuum.

\null

\noindent{\bf B. Dynamical breaking of chiral symmetry}

Making the standard choice of gauge: $A_- =0$~, one finds that
the properties of vector and axial-vector currents (Eqs. (57)-(62)) are also
unaffected by the inclusion of QCD interactions, except for the replacement
of the derivative by the covariant derivative in Eq. (59). The divergence of
the renormalized, space-time, nonsinglet axial current is anomaly-free
(Collins, 1984).
As $j^{\mu\alpha}_{5}$ and $\tilde{j}^{\mu\alpha}_{5}$
become equal in the chiral limit, the divergence of the null-plane current
is also anomaly-free (and goes to zero in the chiral limit).
The corresponding charges, however, {\it do not} become equal
in the chiral limit. This can only be due to contributions at $x^-$-infinity
coming from the Goldstone boson fields, which presumably cancel the
pion pole of the space-time axial charges\footnote{$^1$}{Equivalently, if
one chooses
periodic boundary conditions, one can say that this effect comes from
the longitudinal zero modes of the fundamental fields.}.

{}From soft pion physics we know that the chiral limit of $SU(2)\otimes SU(2)$
is well-described by PCAC. Now, using PCAC one can show that in the chiral
limit $Q^{\alpha}_5~(\alpha=1,2,3)$ is conserved, but $\tilde{Q}^{\alpha}_5$
is not (Ida, 1974; Carlitz {\it et al.}, 1975).
In other words, the renormalized null-plane charges are sensitive to
spontaneous symmetry breaking, although they do annihilate the vacuum.
It is likely that this behavior generalizes to $SU(3)\otimes SU(3)$, viz.,
to the other five lightlike axial charges. Its origin, again, must lie
in terms at infinity/zero modes.

In view of this 'time'-dependence, one might wonder
whether the null-plane axial charges are observables. From PCAC, we know that
it is indeed the case: their
matrix elements between hadron states are directly related to off-shell
pion emission (Feinberg, 1973; Carlitz {\it et al.}, 1975).
For a hadron $A$ decaying into a hadron $B$ and a pion,
one finds
$$
<B|\tilde{Q}^{\alpha}_5 (0)|A>=-{2i(2\pi)^3 p^{+}_A \over
m^{2}_A -m^{2}_B}<B,\pi^{\alpha}|A>\delta^3 (\tilde{p}_A -
\tilde{p}_B)~~. \eqno(75)
$$
Note that in this reaction,
the mass of hadron $A$ must be larger than the mass of $B$  due to the
pion momentum.

\null

\noindent{\bf C. Physical multiplets}

Naturally, we shall assume  that real hadrons fall into representations of
an $SU(3)\otimes SU(3)$ algebra. In Sec. II.E, we have identified
the generators of this algebra with the lightlike chiral charges. But this
was done in the artificial case of the free quark model. It remains to
check whether this identification works in the real world.

Of course, we already know that the predictions based on isospin ($\alpha
=1,2,3$) and hypercharge ($\alpha=8$) are true. Also, the nucleon-octet ratio
$D/F$ is correctly predicted to be 3/2, and several relations between
magnetic moments match well with experimental data.

Unfortunately, several other predictions are in disagreement with observations
(Close, 1979).
For example, $G_A/G_V$ for the nucleon is expected to be equal to 5/3,
while
the experimental value is about 1.25. Dominant decay channels such as $N^*
\rightarrow N\pi$, or $b_1 \rightarrow\omega\pi$, are forbidden by the
lightlike current algebra. The anomalous magnetic moments of nucleons,
and all
form factors of the rho-meson would have to vanish.
De Alwis and Stern (1974) point out that the matrix element
of $\tilde{j}^{\mu\alpha}$ between two given hadrons would be equal to the
matrix element of $\tilde{j}^{\mu\alpha}_5$ between the same two hadrons, up
to a ratio of Clebsch-Gordan coefficients. This is excluded though because
vector and axial-vector form factors have very different analytic
properties as functions of momentum-transfer.

In addition there is, in general, disagreement between the values of $L_z$
assigned to any given hadron. This comes about because in the classification
scheme, the value of $L_z$ is essentially an afterthought, when
group-theoretical considerations based on flavor and helicity have been
taken care of. On the other hand, at the level of the current quarks, this
value is determined by covariance and external symmetries. Consider for
example the $L_z$ assignments in the case of the pion, and of the rho-meson
with zero helicity. As we mentioned earlier, the classification assigns
to these states a pure value of $L_z$, namely zero. However, at the fundamental
level, one expects these mesons to contain a wave-function
$\phi_1$ attached to $L_z =0$ (antiparallel $q\bar q$ helicities), and
also a wave-function $\phi_2$ attached to $L_z =\pm 1$ (parallel
helicities). Actually, the distinction between the pion and the zero-helicity
rho is only based on the different momentum-dependence of $\phi_1$ and $\phi_2$
(Leutwyler, 1974b, 1974c).
If the interactions were turned off, $\phi_2$ would vanish and the
masses of the two mesons would be degenerate (and equal to $(m_u +m_d)$).

We conclude from this comparison with experimental data, that if indeed real
hadrons are representations of some $SU(3)\otimes SU(3)$ algebra, then the
generators $G^{\alpha}$ and $G^{\alpha}_5$
of this classifying algebra must be different from the current lightlike
charges $\tilde{Q}^{\alpha}$ and $\tilde{Q}^{\alpha}_{5}$ (except
however for $\alpha=1,2,3,8$). Furthermore, in order to avoid the
phenomenological discrepancies discussed above, one must forego
kinematical invariance for these generators; that is, $G^{\alpha}(\tilde{k})$
and $G^{\alpha}_5 (\tilde{k})$ must depend on the momentum $\tilde k$ of the
hadrons in a particular irreducible multiplet.

Does that mean that our efforts
to relate the physical properties of hadrons to the underlying field
theory turn out to be fruitless? Fortunately no, as argued by De Alwis
and Stern (1974). The fact that these two sets of generators (the
$\tilde{Q}$'s and the $G$'s) act in the same Hilbert space, in addition to
satisfying  the same
commutation relations, implies that they must actually be unitary equivalent
(this equivalence was originally suggested by Dashen, and by Gell-Mann,
1972a, 1972b). There exists
a set of momentum-dependent unitary operators $U(\tilde {k})$ such that
$$
G^{\alpha}(\tilde {k})=U(\tilde {k})\tilde {Q}^{\alpha}U^{\dagger}(\tilde{k})
{}~,~G^{\alpha}_{5}(\tilde {k})=U(\tilde {k})\tilde {Q}^{\alpha}_{5}
U^{\dagger}(\tilde{k})~~.\eqno (76)
$$
Current quarks, and the real-world hadrons built out of them,
fall into representations of this algebra. Equivalently (e.g., when
calculating electroweak matrix elements), one may
consider the original current algebra, and define its representations
(namely, the ones constructed in Sec. II.E) as
'constituent' quarks and 'constituent' hadrons. These quarks (and antiquarks)
within a hadron of momentum $\tilde {k}\,$ are represented
by a 'constituent fermion field'
$$
\chi ^{\tilde {k}}_{+}(x){\Bigl\vert}_{x^+ =0}\,\equiv U(\tilde {k})\psi_{+}(x)
{\Bigl\vert}_{x^+ =0}~U^{\dagger}(\tilde {k})~~,\eqno (77)
$$
on the basis of which the physical generators
can be written in canonical form:
$$
\tilde {G}^{\alpha} \equiv \int d^3 {\tilde x} \,\bar{\chi} \gamma^{+}
{\lambda^{\alpha}\over 2}\chi~,~
\tilde {G}^{\alpha}_5  \equiv \int d^3 {\tilde x} \,\bar{\chi} \gamma^{+}
\gamma_5 {\lambda^{\alpha}\over 2}\chi~~. \eqno (78)
$$ From Eq. (77), it follows that the constituent annihilation/creation
operators are derived from
the current operators via
$$
a^{\tilde {k}}\,(\tilde {p},h)\equiv U(\tilde {k})b (\tilde {p},h)
U^{\dagger}(\tilde {k})~~,~~{c^{\tilde {k}}\,}^{\dagger}(\tilde {p},h)
\equiv
U(\tilde {k})d^{\dagger}(\tilde {p},h)U^{\dagger}(\tilde {k})~~.
\eqno (79)
$$

Due to isospin invariance, this unitary transformation cannot mix flavors,
it only mixes helicities. It can therefore be represented by three unitary
$2\times 2$ matrices $T^{f}(\tilde {k},\tilde {p})$ such that
$$
a^{\tilde {k}}_{f}(\tilde {p},h)=\sum_{h'=\pm{1\over 2}}
T^{f}_{hh'}(\tilde {k},\tilde {p})\,b_f
(\tilde {p},h')~,~c^{\tilde {k}}_{f}(\tilde {p},h)=\sum_{h'=\pm{1\over 2}}
T^{f\,*}_{hh'}(\tilde {k},\tilde {p})\,d_f (\tilde {p},h')~~, \eqno (80)
$$
one for each flavor $f=u,d,s$.
Since we need the transformation to be unaffected when
$\tilde {k}$ and $\tilde {p}$ are boosted along $z$
or rotated around $z$ together,
the matrix $T$ must actually be a
function of only kinematical invariants. These are
$$
\xi \equiv {p^+ \over {k^+}}~~{\rm and}~~{\bf\kappa}_{\perp}\equiv
{\bf p}_{\perp}-\xi{\bf k}_{\perp}~~,~~{\rm where}
\sum_{\rm constituents}\xi =1~~,~~\sum_{\rm constituents}{\bf\kappa}_{\perp}
={\bf 0}~~.\eqno (81)
$$
Invariance under time reversal ($x^+ \mapsto -x^+$) and parity ($x^1
\mapsto -x^1$) further constrain its functional form, so that finally
$$
T^{f}(\tilde {k},\tilde {p})=\exp\,[-i\,{{\bf\kappa}_{\perp}\over |{\bf\kappa}
_{\perp}|}\cdot{\bf\sigma}_{\perp}~\beta _{f}(\xi,{\bf\kappa}_{\perp}^{2})]
\eqno (82)
$$
(Leutwyler, 1974b, 1974c).

Thus the relationship between current and constituent quarks is embodied
in the three functions $\beta _f (\xi,{\bf\kappa}_{\perp}^{2})$, which we
must try to extract from comparison with experiment. (In first approximation
it is legitimate to take $\beta _u$ and $\beta _d$ equal since $SU(2)$
is such a  good symmetry.)

Based on some assumptions abstracted from the free-quark model (as suggested
by Fritzsch and Gell-Mann, 1972), Leutwyler (1974a, 1974b, 1974c)
has derived a set of sum rules
obeyed by mesonic wave-functions. Implementing then the transformation
described
above, Leutwyler finds various relations involving form factors and scaling
functions of mesons, and computes the current quark masses. For example, he
obtains
$$
F_{\pi}<F_{\rho}~~,~~F_{\rho}=3\,F_{\omega}~~,~~F_{\rho}<{3\over {\sqrt 2}}\,
|F_{\phi}|~~, \eqno (83)
$$
and the $\omega /\phi$ mixing angle is estimated to be about 0.07 rad.
Leutwyler (1974a) also shows
that the average transverse momentum of a quark inside a meson is substantial
($|{\bf p}_{\perp}|_{\rm rms} > 400$ MeV), thus justifying
{\it a posteriori} the basic assumptions of the relativistic CQM (e.g.,  Fock
space truncation and relativistic energies). This large value also provides
an explanation for the above-mentioned failures of the
$SU(3)\otimes SU(3)$ classification scheme (Close, 1979).

On the negative side, it appears that the functional dependance of
the $\beta _f$'s cannot be easily determined with satisfactory precision.

\null

\noindent{\bf IV. SUMMARY AND OUTLOOK}

In this review, we have studied the properties of vector and axial-vector
nonsinglet charges, and compared their space-time with their null-plane
realization. We have shown that the free-quark model in a null-plane
frame is chirally symmetric in the $SU(3)$ limit, whether the common
mass is zero or not. The difference between space-time and null-plane
chiral properties clearly shows
up at the level of the axial currents; this feature has not been exhibited
before this work.

In QCD, chiral symmetry is broken both explicitly and dynamically. This
is reflected in a null-plane frame by the fact that the axial charges are
not conserved, even in the 'chiral limit'. Vector and axial-vector
charges annihilate the Fock vacuum, and so are {\it bona fide} operators.
They form an $SU(3)\otimes SU(3)$ algebra, and conserve the number of
quarks and anti-quarks separately when acting on a hadron state. Hence they
classify hadrons, on the basis of their valence structure, into multiplets,
which are not mass-degenerate.

This classification however turns out to be phenomenologically deficient.
The remedy to this situation is a unitary transformation between the
charges and the (postulated) physical generators of the classifying
$SU(3)\otimes SU(3)$ algebra. The latter generators are built out of
'constituent' quark fields produced by the unitary transformation from the
fundamental 'current' fields. The functional form of this transformation
can be strongly constrained based on symmetry arguments. Since it depends
necessarily on the momenta of the hadron multiplets, the transformation
does not commute with the null-plane
'energy', and the constituent 'masses' are themselves momentum-dependent.
Given that the CQM's have been quite successful while using of course
fixed masses, one expects that the variation of constituent masses over a
typical
range of momenta might be small compared to their average value.

That quark masses are allowed to vary, is not surprising in view of the fact
that the constituent quarks are not elementary particles, but composite
systems containing current quarks of both helicities, as displayed in
Eq. (80). We observe that the null-plane picture of a constituent
quark (of a given flavor) is much simpler than the space-time description,
since the latter must include both current quarks and current antiquarks,
of all three flavors, in addition to the expected spin mixing (Weinberg, 1990;
Fritzsch, 1993).

Further research is needed on several key aspects of the approach to
hadron phenomenology which has been reviewed in this paper.
In Sec. III.C, we have discussed applications
to the meson sector; but these methods
have not yet (to our knowledge) been applied to baryons. It would also be
gratifying if one could devise a calculation
of constituent masses {\it within this framework}, and find the correct
order of magnitude.
Such a derivation could, by the way, only benefit from a better knowledge
of the
functions $\beta_f
(\xi,{\bf\kappa}_{\perp}^{2})$ which parametrize the relationship
between current and constituent quarks.

\null

\noindent{\bf ACKNOWLEDGMENTS}

We would like to thank Stan Brodsky, Avaroth Harindranath, and Ken
Wilson for illuminating discussions, and Steve Pinsky for continuous
encouragement and support.

\null

\noindent{\bf APPENDIX: BOUNDARY TERM FOR FREE-FERMION NUMBER}

The system is confined in a 'box' whose boundaries extend from: $z=-L
\sqrt{2}$ to $z=+L\sqrt{2}$ at $t=0$, from $x^- =-L$ to $x^- =+L$ at
$x^+ =0$, from $x^+ =-L$ to $x^+ =0$ at $x^- =+L$, and from $x^+ =0$ to
$x^+ =+L$ at $x^- =-L$.

We are concerned here with the last two hypersurfaces. We wish to prove
that
$$
B\equiv \lim_{L\to\infty} \int d^2 {\bf x}_{\perp} \Biggl[ \int_{-L}^{0}
dx^+ \,j^- (x){\Bigl\vert}_{x^- =+L}+\int_{0}^{L} dx^+ \,j^- (x)
{\Bigl\vert}_{x^- =-L}\Biggr]~~,\eqno (A1)
$$
(where~~~$j^- =\bar\psi \gamma^- \psi$) is zero.
\medskip
First, we insert in this expression the momentum-space expansion of the
fermion field. The system is confined in a longitudinal 'box', hence
$p^+$ is discretized. However, since we are going to take the limit $L\to
\infty$ at the end of this calculation, we may use right away the
infinite-volume expansion. Using Eq. (15) and Eqs. (43)-(45), one finds
$$
\psi(x)=\int{d^2{\bf p}_\perp\over(2\pi)^{3/2}}\int_{0}^{+\infty}\!
{dp^+\over\sqrt{2p^+}}
\sum_{h=\pm {1\over 2}}\Biggl[u(\tilde{p},h)e^{-ipx} b(\tilde{p},h)
+v(\tilde{p},h)e^{+ipx}d^{\dagger}(\tilde{p},h)\Biggr]~~,\eqno (A2)
$$
where the creation and annihilation operators are the same as in Eq. (44),
and
$$
\sum_{h=\pm {1\over 2}}u(\tilde{p},h)\bar u (\tilde{p},h)=\not\!p +m~,~
\sum_{h=\pm {1\over 2}}v(\tilde{p},h)\bar v (\tilde{p},h)=\not\!p -m~~.
\eqno (A3)
$$

Inserting Eq. (A2) in the expression of the current yields
$$\eqalign{
&j^- (x){\Bigl\vert}_{x^- =\Lambda}=\int{d^2{\bf p}_\perp d^2{\bf q}_{\perp}
\over(2\pi)^3}\int_{0}^{+\infty}\!{dp^+ dq^+\over 2\sqrt{p^+ q^+}}
\sum_{h,\eta =\pm {1\over 2}}\cr
&~\Biggl[e^{+iq^+ \Lambda}e^{+i(q^- x^+ -{\bf q}_{\perp}\cdot
{\bf x}_{\perp})} \bar{u}(\tilde{q},\eta) b^{\dagger}(\tilde{q},\eta)
+e^{-iq^+ \Lambda}e^{-i(q^- x^+ -{\bf q}_{\perp}\cdot
{\bf x}_{\perp})} \bar{v}(\tilde{q},\eta)d(\tilde{q},\eta)\Biggr]~\gamma^-\cr
&~\Biggl[e^{-ip^+ \Lambda}e^{-i(p^- x^+ -{\bf p}_{\perp}\cdot
{\bf x}_{\perp})} u(\tilde{p},h) b(\tilde{p},h)
+e^{+ip^+ \Lambda}e^{+i(p^- x^+ -{\bf p}_{\perp}\cdot
{\bf x}_{\perp})}\!v(\tilde{p},h)d^{\dagger}(\tilde{p},h)\!\Biggr]~~~.}
\eqno (A4)
$$
Integrating over ${\bf x}_{\perp}$, and making use of the spinor identities
$$\eqalign{
\bar{u}(\tilde{q},\eta)\gamma^- u(\tilde{p},h)
{\Bigl\vert}_{{\bf q}_{\perp}={\bf p}_{\perp}}
=2p^- &\sqrt{p^+ \over q^+}~\delta_{h\eta}
=\bar{v}(\tilde{q},\eta)\gamma^- v(\tilde{p},h){\Bigl\vert}_{{\bf q}_{\perp}=
{\bf p}_{\perp}}~~,\cr
\bar{u}(\tilde{q},\eta)\gamma^- v(\tilde{p},h){\Bigl\vert}_{{\bf q}_{\perp}=
-{\bf p}_{\perp}}
=-2p^- &\sqrt{p^+ \over q^+}~\delta_{h,-\eta}
=\bar{v}(\tilde{q},\eta)\gamma^- u(\tilde{p},h){\Bigl\vert}_{{\bf q}_{\perp}=
-{\bf p}_{\perp}}~~,}\eqno (A5)
$$
one gets
$$\eqalign{
&\int d^2 {\bf x}_{\perp}\,j^- (x){\Bigl\vert}_{x^- =\Lambda}=\int
{d^2{\bf p}_\perp
\over 2\pi}\int_{0}^{+\infty}\! dp^+ dq^+ \Bigl({p^- \over q^+}\Bigr)
\sum_{h=\pm {1\over 2}}\cr
&~\Biggl[e^{i\Lambda(q^+ -p^+)}e^{ix^+ (q^- -p^-)}b^{\dagger}(\tilde{q},h)
b(\tilde{p},h)+e^{-i\Lambda(q^+ -p^+)}e^{-ix^+ (q^- -p^-)}d(\tilde{q},h)
d^{\dagger}(\tilde{p},h)~~~~~~~(A6)\cr
&~~-\!e^{i\Lambda(q^+ +p^+)}e^{ix^+ (q^- +p^-)}b^{\dagger}(\tilde{q}',-h)
d^{\dagger}(\tilde{p},h)
\!-\!e^{-i\Lambda(q^+ +p^+)}e^{-ix^+ (q^- +p^-)}d(\tilde{q}',-h)
b(\tilde{p},h)\Biggr]~,}
$$
where ~~$\tilde{q}\equiv (q^+,{\bf p}_{\perp})$, and ~~$\tilde{q}'\equiv
(q^+,-{\bf p}_{\perp})$~. Inserting Eq. (A6) into (A1), one obtains
$$\eqalign{
B=\int{d^2{\bf p}_\perp
\over 2\pi}\int_{0}^{+\infty}\! dp^+ dq^+ &\Bigl({p^- \over q^+}\Bigr)
\sum_{h=\pm {1\over 2}}
\Biggl[T(\tilde{p},\tilde{q})b^{\dagger}(\tilde{q},h)b(\tilde{p},h)\!+
\!T^{*}(\tilde{p},\tilde{q})d(\tilde{q},h)d^{\dagger}(\tilde{p},h)\cr
&-\!T'(\tilde{p},\tilde{q}') b^{\dagger}(\tilde{q}',-h)d^{\dagger}
(\tilde{p},h)
\!-\!T'^{*}(\tilde{p},\tilde{q})d(\tilde{q}',-h)b(\tilde{p},h)\Biggr],~~(A7)}
$$
where
$$\eqalign{
T(\tilde{p},\tilde{q}) &\equiv \lim_{L\to\infty}\biggl[e^{+iL(q^+ -p^+)}
\!\int_{-L}^{0} dx^+
e^{ix^+ (q^- -p^-)}\,+\,e^{-iL(q^+ -p^+)}
\!\int_{0}^{L} dx^+ e^{ix^+ (q^- -p^-)}\biggr]~,\cr
T'(\tilde{p},\tilde{q}') &\equiv \lim_{L\to\infty}\biggl[e^{+iL(q^+ +p^+)}
\!\int_{-L}^{0} dx^+
e^{ix^+ (q^- +p^-)}\,+\,e^{-iL(q^+ +p^+)}
\!\int_{0}^{L} dx^+ e^{ix^+ (q^- +p^-)}\biggr]~.}\eqno (A8)
$$
\medskip
Next, we show that both $T$ and $T'$ are identically zero, thus
obviously making
$B$ zero, as advertised. Consider the first term within the square brackets
in the expression of $T$:
$$\eqalign{
&e^{+iL(q^+ -p^+)}\!\int_{-L}^{0} dx^+ e^{ix^+ (q^- -p^-)}=
e^{+iL(q^+ -p^+)}~{1-e^{-iL(q^- -p^-)}\over i(q^- -p^-)}\cr
&={1\over i(q^- -p^-)}\Bigl[\bigl(e^{iL(q^+ -p^+)}-1)-
(e^{iL(q^+ -p^+ -q^- +p^-)}-1\bigr)\Bigr]\cr
&={1\over i(q^- \!-\!p^-)}\biggl[i(q^+ \!-\!p^+)\!\int_{0}^{L}\!dy\,
e^{iy(q^+ -p^+)}
-i(q^+ \!-\!p^+ \!-\!q^- \!+\!p^-)\!\int_{0}^{L}\!dy\,
e^{iy(q^+ -p^+ -q^- +p^-)}\biggr]\cr
&=-{q^+ \over p^-}\int_{0}^{L}dy~e^{iy(q^+ -p^+)}
+\Bigl(1+{q^+ \over p^-}\Bigr)\int_{0}^{L}
dy~e^{iy(q^+ -p^+ -q^- +p^-)}~~.~~~~~~~~~~~~~~~~~~~~~~~~(A9)}
$$
In the last step, we exploited the fact that
$$
q^- -p^- =p^- \,\biggl[{q^- \over p^-}-1\biggr]=p^- \Biggl[{\bigl(
{{\bf p}_{\perp}^2 +m^2 \over 2q^+}\bigr)
\over\bigl({{\bf p}_{\perp}^2 +m^2 \over 2p^+}\bigr)}-1\Biggr]=
p^- \,\Bigl({p^+ \over q^+}-1\Bigr)={p^- \over q^+}\,(p^+ -q^+)~~.
\eqno (A10)
$$
Similarly, the second term in $T$ is
$$\eqalign{
&e^{-iL(q^+ -p^+)}\!\int_{0}^{L} dx^+ e^{ix^+ (q^- -p^-)}\cr
&~=
-{q^+ \over p^-}\int_{-L}^{0}dy~e^{iy(q^+ -p^+)} +\Bigl(1+{q^+ \over p^-}
\Bigr)\int_{-L}^{0}dy~e^{iy(q^+ -p^+ -q^- +p^-)}~~.}\eqno (A11)
$$
Putting Eqs. (A9) and (A11) into Eq. (A8) yields
$$
T=-{q^+ \over p^-}\int_{-\infty}^{\infty}dy~e^{iy(q^+ -p^+)} +
\Bigl(1+{q^+ \over p^-}\Bigr)
\int_{-\infty}^{\infty}dy~e^{iy(q^+ -p^+ -q^- +p^-)}~~.\eqno (A12)
$$
Finally,
$$\eqalign{
T &=-{q^+ \over p^-}\int_{-\infty}^{\infty}dy~e^{iy(q^+ -p^+)} +
{q^+ \over p^-}\Bigl(1+{p^- \over q^+}\Bigr)
\int_{-\infty}^{\infty}dy~e^{iy(1+{p^- \over q^+})(q^+ -p^+)}\cr
&=-{q^+ \over p^-}\int_{-\infty}^{\infty}dy~e^{iy(q^+ -p^+)} +
{q^+ \over p^-}
\int_{-\infty}^{\infty}du~e^{iu(q^+ -p^+)}=0~~.}\eqno (A13)
$$

A similar massage on $T'$ gives
$$
T' ={q^+ \over p^-}\int_{-\infty}^{\infty}dy~e^{iy(q^+ +p^+)} +
\Bigl(1-{q^+ \over p^-}\Bigr)
\int_{-\infty}^{\infty}dy~e^{iy(q^+ +p^+ -q^- -p^-)}\eqno (A14)
$$
analogous to Eq. (A12). Since $(q^+ + p^+)$ is always positive, the first
integral does not contribute, and
we are left with
$$\eqalign{
T' &=-{q^+ \over p^-}\Bigl(1-{p^- \over q^+}\Bigr)
\int_{-\infty}^{\infty}dy~e^{iy(q^+ +p^+)(1-{p^- \over q^+})}\cr
&=-{q^+ \over p^-}\Bigl(1-{p^- \over q^+}\Bigr)
{1\over q^+ +p^+}\int_{-\infty}^{\infty}dz~e^{iz(1-{p^- \over q^+})}\cr
&=-{2\pi q^+\over p^- \,(q^+ +p^+)}\Bigl(1-{p^- \over q^+}\Bigr)
\delta\Bigl(1-{p^- \over q^+}\Bigr)~~,}\eqno (A15)
$$
which is zero as claimed.

\null

\noindent{\bf REFERENCES}

\noindent Bell, J.S., 1974, Acta Phys. Austriaca, Suppl. {\bf 13}, 395.

\noindent Buccella, F., E. Celeghin, H. Kleinert, C.A. Savoy and
E. Sorace, 1970, Nuovo Cimento A

\noindent\enskip {\bf 69}, 133.

\noindent Carlitz, R., D. Heckathorn, J. Kaur and W.-K. Tung, 1975,
Phys. Rev. D {\bf 11}, 1234.

\noindent Carlitz, R., and W.-K. Tung, 1976, Phys. Rev. D {\bf 13}, 3446.

\noindent Collins, J.C., 1984, {\it Renormalization} (Cambridge U.P.,
New York).

\noindent Close, F., 1979, {\it An Introduction to Quarks and Partons}
(Academic, New York).

\noindent De Alwis, S.P., 1973, Nucl. Phys. B {\bf 55}, 427.

\noindent De Alwis, S.P., and J. Stern, 1974, Nucl. Phys. B {\bf 77}, 509.

\noindent Eichten, E., F. Feinberg and J.F. Willemsen, 1973, Phys. Rev. D
{\bf 8}, 1204.

\noindent Feinberg, F.L., 1973, Phys. Rev. D {\bf 7}, 540.

\noindent Fritzsch, H., 1993, ``Constituent Quarks,
Chiral Symmetry and the Nucleon Spin'', talk

\noindent\enskip given at the Sep. 1993
Leipzig Workshop on Quantum Field Theory Aspects of High

\noindent\enskip Energy Physics (CERN preprint TH. 7079/93).

\noindent Fritzsch, H., and M. Gell-Mann, 1972, in {\it Proc. 16th. Int.
Conf. on HEP, Batavia, IL}.

\noindent Gasser, A., and H. Leutwyler, 1975, Nucl. Phys. B {\bf 94},
269.

\noindent Gell-Mann, M., 1972a, Lectures given at the 1972
Schladming Winter School (CERN

\noindent\enskip preprint TH. 1543/72).

\noindent Gell-Mann, M., 1972b, Acta Phys. Austriaca, Suppl. {\bf 9}, 733.

\noindent Georgi, H., 1982, {\it Lie Algebras in Particle Physics}
(Benjamin-Cummings, Reading, M.A.).

\noindent G{\l}azek, S.D., A. Harindranath, R.J. Perry, T. Walhout,
K.G. Wilson, and W.-M. Zhang,

\noindent\enskip 1994, ``Nonperturbative QCD: A
Weak-Coupling Treatment on the Light-Front'', Ohio

\noindent\enskip State Univ. preprint.

\noindent Hasenfratz, P., and J. Kuti, 1978, Phys. Rep. C {\bf 40}, 75.

\noindent Ida, M., 1974, Progr. Theor. Phys. {\bf 51}, 1521.

\noindent Ida, M., 1975a, Progr. Theor. Phys. {\bf 54}, 1199.

\noindent Ida, M., 1975b, Progr. Theor. Phys. {\bf 54}, 1519.

\noindent Ida, M., 1975c, Progr. Theor. Phys. {\bf 54}, 1775.

\noindent Jers\'ak, J., and J. Stern, 1968, Nucl. Phys. B {\bf 7}, 413.

\noindent Jers\'ak, J., and J. Stern, 1969, Nuovo Cimento {\bf 59}, 315.

\noindent Leutwyler, H., 1968, Acta Phys. Austriaca, Suppl. {\bf 5}, 320.

\noindent Leutwyler, H., 1969, in Springer Tracts in Modern Physics, No. 50,
(Springer, New York),

\noindent\enskip p. 29.

\noindent Leutwyler, H., 1974a, Phys. Lett. B {\bf 48}, 45.

\noindent Leutwyler, H., 1974b, Phys. Lett. B {\bf 48}, 431.

\noindent Leutwyler, H., 1974c, Nucl. Phys. B {\bf 76}, 413.

\noindent McCartor, G., 1988, Z. Phys. C {\bf 41}, 271.

\noindent Melosh, H.J., 1974, Phys. Rev. D {\bf 9}, 1095.

\noindent Osborn, H., 1974, Nucl. Phys. B {\bf 80}, 90.

\noindent Sazdjian H., and J. Stern, 1975, Nucl. Phys. B {\bf 94}, 163.

\noindent Shifman, M.A., A.I. Vainshtein and V.I. Zakharov, 1979,
Nucl. Phys. B {\bf 147}, 385.

\noindent Weinberg, S., 1990, Phys. Rev. Lett. {\bf 65}, 1181.

\bye